\documentclass[twocolumn,aps,prl,showpacs,superscriptaddress]{revtex4}

\usepackage{graphicx}

\def\bra#1{\left\langle#1\right|}
\def\ket#1{\left|#1\right\rangle}
\def\ketbra#1#2{\left|#1\right\rangle\!\left\langle#2\right|}

\def\captionstyle{\em}

\begin{document}

\title{Quantum non-Gaussian Depth of Single-Photon States}

\author{Ivo Straka}
\email{straka@optics.upol.cz}
\affiliation{Department of Optics, Palack\' y University, 17. listopadu 1192/12,  771~46 Olomouc, Czech Republic}

\author{Ana Predojevi\' c}
\affiliation{Institut f\" ur Experimentalphysik, Universit\" at Innsbruck, Technikerstra\ss e 25, 6020 Innsbruck, Austria}

\author{Tobias Huber}
\affiliation{Institut f\" ur Experimentalphysik, Universit\" at Innsbruck, Technikerstra\ss e 25, 6020 Innsbruck, Austria}

\author{Luk\' a\v s Lachman}
\affiliation{Department of Optics, Palack\' y University, 17. listopadu 1192/12,  771~46 Olomouc, Czech Republic}

\author{Lorenz Butschek}
\affiliation{Institut f\" ur Experimentalphysik, Universit\" at Innsbruck, Technikerstra\ss e 25, 6020 Innsbruck, Austria}

\author{Martina Mikov\' a}
\affiliation{Department of Optics, Palack\' y University, 17. listopadu 1192/12,  771~46 Olomouc, Czech Republic}

\author{Michal Mi\v cuda}
\affiliation{Department of Optics, Palack\' y University, 17. listopadu 1192/12,  771~46 Olomouc, Czech Republic}

\author{Glenn S. Solomon}
\affiliation{Joint Quantum Institute, National Institute of Standards and Technology, and University of Maryland, Gaithersburg, Maryland 20849, USA}

\author{Gregor Weihs}
\affiliation{Institut f\" ur Experimentalphysik, Universit\" at Innsbruck, Technikerstra\ss e 25, 6020 Innsbruck, Austria}

\author{Miroslav Je\v zek}
\affiliation{Department of Optics, Palack\' y University, 17. listopadu 1192/12,  771~46 Olomouc, Czech Republic}

\author{Radim Filip}
\affiliation{Department of Optics, Palack\' y University, 17. listopadu 1192/12,  771~46 Olomouc, Czech Republic}

\begin{abstract}
We introduce and experimentally explore the concept of quantum non-Gaussian depth of single-photon states with a positive Wigner function. The depth measures the robustness of a single-photon state against optical losses. The directly witnessed quantum non-Gaussianity withstands significant attenuation, exhibiting a depth of 18 dB, while the nonclassicality remains unchanged. Quantum non-Gaussian depth is an experimentally approachable quantity that is much more robust than the negativity of the Wigner function. Furthermore, we use it to reveal significant differences between otherwise strongly nonclassical single-photon sources.
\end{abstract}


\maketitle

\begin{figure}[b]

	\includegraphics[width=.9\columnwidth]{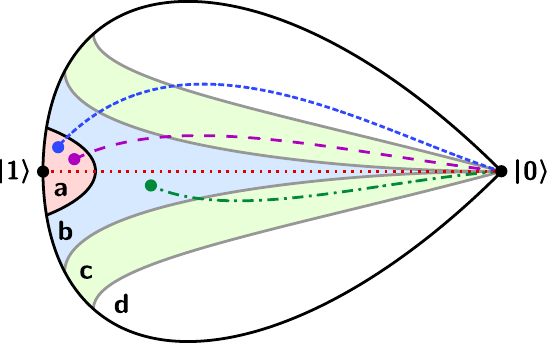}
	\caption[Nonclassical sets]{The classification of quantum state sets used in our discussion, where $\textsf{\textbf{a}} \subset \textsf{\textbf{b}} \subset \textsf{\textbf{c}} \subset \textsf{\textbf{d}}$. \textsf{\textbf{a}} is the set of states with negative Wigner function. All states in the set \textsf{\textbf{b}} are guaranteed to be quantum non-Gaussian. Likewise, all states in the set \textsf{\textbf{c}} are nonclassical. The set \textsf{\textbf{d}} contains all states in general. Equivalently, all classical states are contained in the complement $\bar\textsf{\textbf{c}}$ and all Gaussian mixtures are in $\bar\textsf{\textbf{b}}$. The borders of \textsf{\textbf{b}} and \textsf{\textbf{c}} therefore represent NC and QNG witnesses.

	The points and non-solid lines represent various realistic quantum states and their respective paths under attenuation. These quantum states can approach vacuum inside different sets. An ideal single-photon state ($P_0\ketbra{0}{0}+P_1\ketbra{1}{1}$, red dotted line) is an extremal case of an infinitely robust QNG state. Other realistic states may exhibit infinite NC depth or leave nonclassical states altogether (dashed lines). The green dot-dashed line represents realistic single-photon states with positive Wigner function, as generated experimentally.
	}
	\label{fig.sets}
	
\end{figure}

{\captionstyle Introduction.---}
Single-photon states are important resources used in quantum computation and information processing \cite{book}. Furthermore, they are direct evidence of the quantum nature of light \cite{Einstein}. This nature is manifested via various quantum features of the single-photon state. Some of these features rely on distinguishing the single-photon state from statistical mixtures of certain quantum states, such as coherent or Gaussian states. Such convex sets then serve to define the respective quantum features of a single photon. Specifically, nonclassicality means the state is inexpressible as a statistical mixture of classical coherent states \cite{Glauber,Mandel}. In addition to this, the state may be inexpressible as a mixture of pure Gaussian states, thus exhibiting quantum non-Gaussianity \cite{Filip,Jezek} (note the difference from classical non-Gaussianity \cite{classnong1,classnong2}). Even further, the state can have a negative Wigner function \cite{Wigner,Lvovsky}. These quantum features define three sets of quantum states, each being a subset of the previous one (see Fig. \ref{fig.sets}).  Depending on its qualities, a realistic single-photon source will produce a state that can belong to any of these subsets. In such a case, the necessary and sufficient condition for nonclassicality is represented by an infinite hierarchy of criteria \cite{Vogel1,Vogel2}.

Here we perform a conceptual evaluation of quantum features of light. We use the important fact that any optical application inevitably includes losses. We do not, however, examine the role of losses in a \emph{specific} application. Instead, we aim to test how well an experimentally generated single-photon state keeps the properties of an ideal single photon. In this protocol-independent approach, the endurance of quantum features with respect to losses becomes imperative \cite{Ursin,Fedrizzi,Grangier2,Yoshikawa}.

The nonclassicality depth is defined as the maximum attenuation of a nonclassical state, at which the state is still able to preserve the nonclassicality \cite{depth1,depth2}. To determine this depth based on its definition, one requires homodyne tomography \cite{Lvovsky} and quantum estimation of the entire density matrix of all emitted modes of light \cite{Rehacek}. In practice, such a measurement is only feasible for a fixed low number of photons in a few well-defined modes. However, many experimentally generated single-photon states exhibit a complex multimode structure. Moreover, they are influenced by multiphoton contributions, which are systematically generated and/or coupled from the environment as noise. Although these contributions alone might not destroy the quantum features of the generated state, they will affect the state and can substantially limit the depth of the respective feature. In Ref. \cite{Jezek}, increased detection noise was simulated and the results show that it can indeed have a destructive effect on quantum non-Gaussianity. Therefore, noise is a major limiting factor and multiphoton contributions need to be taken into account. Unfortunately, such contributions are commonly hard to fully estimate and characterize.

As alternatives to full quantum state estimation, directly measurable witnesses can detect the quantum features of multimode states. The witnesses of nonclassicality (NC) \cite{Glauber,Mandel} and quantum non-Gaussianity (QNG) \cite{Filip,Jezek,nongauss1,nongauss2,nongauss3,nongauss4} allow one to experimentally determine a lower bound on their respective depths. In this Letter, we show a direct measurement of the lower bound on the depths of both NC and QNG for three different sources of single-photon states: two are based on spontaneous parametric down-conversion in a nonlinear crystal, and one is a single quantum dot. Using this approach, we compare the quantum features of these very different single-photon sources. Our measurements show an extreme robustness of NC for single-photon states with positive Wigner function. Our results further prove that QNG is a robust resource for future quantum applications of single-photon states, as opposed to the fragile negativity of the Wigner function. We demonstrate preserving the QNG of a single-photon state for up to 18~dB of attenuation.

{\captionstyle NC and QNG witnesses.---}
Both NC and QNG of a quantum state $\rho$ can be recognized using the criteria derived in Refs. \cite{Filip,Jezek,hierarchy}, which are defined using the probabilities $P_0=\bra{0}\rho\ket{0}$ and $P_1=\bra{1}\rho\ket{1}$. Reference~\cite{hierarchy} shows that a sufficient condition for quantum nonclassical states is that $P_1>-P_0\ln P_0$. Since the error probability of multiphoton contributions $P_{2+}=1-P_0-P_1$, the NC and QNG criteria can be rewritten in terms of $P_1$ and $P_{2+}$. Complete knowledge of the photon statistics is not required, because $P_1$ and $P_{2+}$ are sufficiently informative parameters describing the main statistical properties of light emitted by a single-photon source. These parameters can be efficiently estimated \cite{Jezek} using an autocorrelation measurement \cite{Mandel,Grangier}. An alternative approach, directly extending the anticorrelation parameter \cite{Grangier} traditionally used for the characterization of NC, has been recently proposed in Refs. \cite{hierarchy,Lachman}.

\begin{figure}[b]

			\includegraphics[width=\columnwidth]{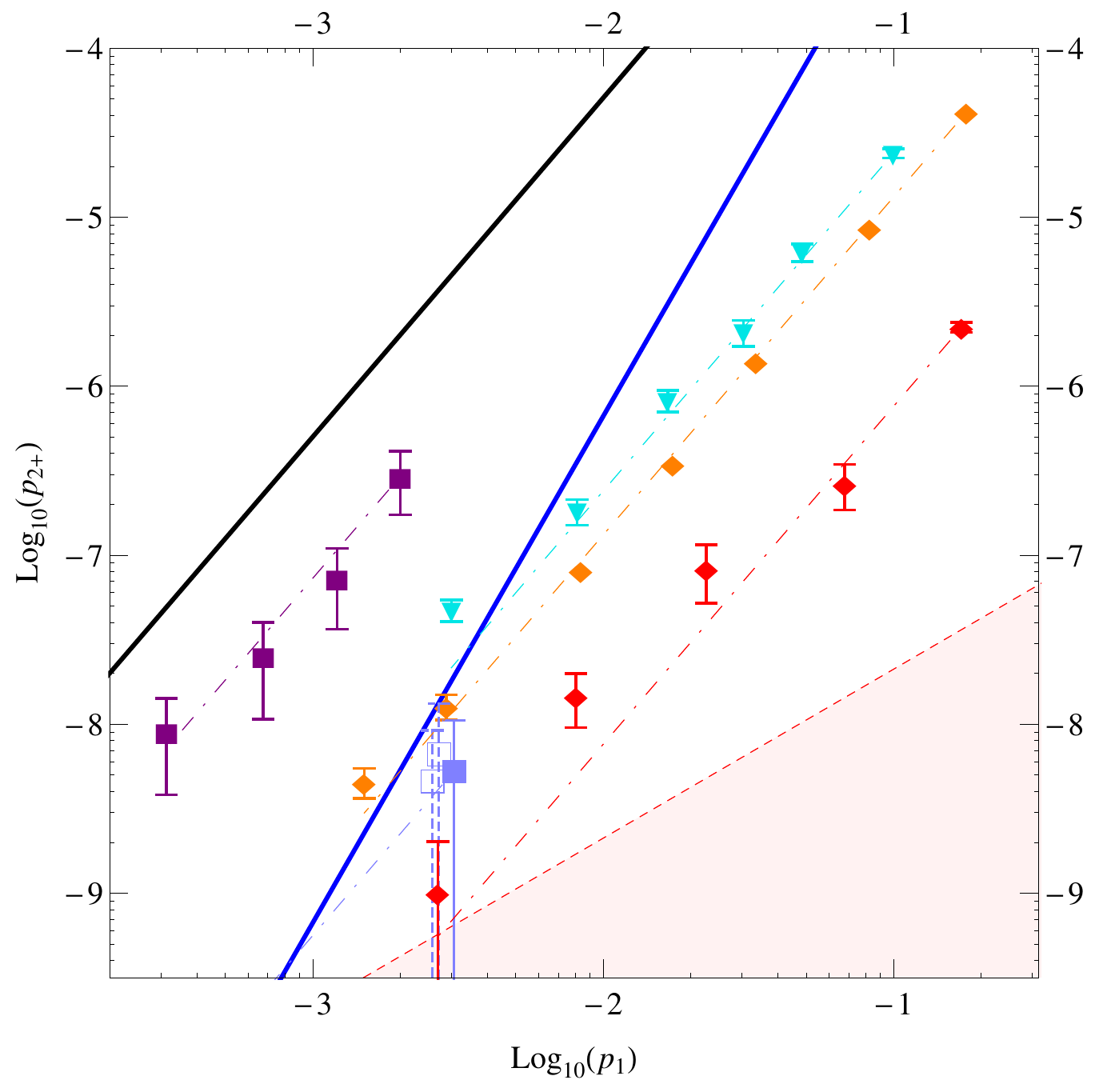}
			\caption[Estimated probabilities]{Estimated probabilities of heralded single-photon states, each series representing various attenuations of a particular state. Full diamonds denote the cw SPDC source: orange ---coincidence window 2 ns; red---low pump, coincidence window 2 ns.
			Cyan triangles denote the pulsed SPDC source.
			Square markers denote the quantum dot: purple squares---above-band excitation, cw pump; blue squares---resonant excitation, pulsed pump.
			The dot-dashed lines represent theoretical prediction of attenuation from the initial point. The dashed red line is the limit of dark counts for the red attenuation data.
			There are two lower bounds for $p_{2+}$; the solid black line is a bound for classical states and the blue line is for Gaussian mixtures. Quantum states below these lines are NC or QNG, respectively.
			Error bars are determined by error propagation from event count errors \cite{Jezek}. Horizontal error bars are smaller than plot points.
			}
			\label{fig.mainplot}

\end{figure}

{\captionstyle Depth of quantum features.---} The lower bound on the depth of a quantum feature of a quantum state can be operationally defined as the maximal attenuation at which that quantum feature is still detectable. We propose physical variable attenuation in front of an autocorrelation measurement and application of the criteria for NC and QNG on the measured data, which is discussed in detail in \cite{Jezek,hierarchy} and plotted in Fig.~\ref{fig.mainplot}. The attenuation of the generated state transforms the photon-number statistics $P_n$ to $P'_n=\sum_{m=n}^{\infty}{m \choose n} T^{n}(1-T)^{m-n}P_m$, where $T$ is the variable transmittance of the attenuator. Because $P_n$ before the attenuation cannot be estimated by the autocorrelation measurement, we implement the attenuation experimentally to determine the depth.

Let us assume an ideal single-photon state $\tilde\rho=\eta \ketbra{1}{1}+(1-\eta)\ketbra{0}{0}$, influenced solely by attenuation. It is straightforward to show that such a state exhibits {\em infinite} NC and QNG depths for any $\eta>0$, while the negativity of the Wigner function vanishes for $\eta\leq 0.5$.
In reality, presence of multiphoton contributions can substantially limit the depths of QNG  and even NC (see Fig. \ref{fig.sets}) \cite{Lachman}. The proposed detection scheme is able to distinguish these realistic single photon states from the state $\tilde\rho$.

{\captionstyle Robust QNG as a single-photon benchmark.---} For a single-photon source of sufficiently high quality, the generated states can exhibit a remarkably large QNG depth despite their multimode background noise, as has been very recently predicted in Ref.~\cite{Lachman}. Then, these realistic states are highly robust single-photon states, approaching the ideal state $\tilde\rho$. Such states are typically generated by high-quality single-photon sources, where the generated state can be very well approximated by a mixture
\begin{equation}
	\label{eq.hqstate}
	\rho\approx  (1-P_1-P_{2+})\ketbra{0}{0}+P_1\ketbra{1}{1}+P_{2+}\ketbra{2}{2},
\end{equation}
where $P_{2+}\ll P_1$ and we do not distinguish between photons in different modes. Using the parametrization of $P_1$ and $P_{2+}$, the criterion for NC is approximately given by $P_{2+}<\frac{1}{2}P_1^2$, whereas the criterion for QNG can be approximated by $P_{2+}<\frac{2}{3}P_1^3$ \cite{Lachman}. After the attenuation, we neglect the transfer from the state $\ketbra{2}{2}$ to $\ketbra{1}{1}$, simplifying our description. Explicitly, we use a lower bound $TP_1$ on $P'_1\geq TP_1$, which is safe from false QNG witnessing, and $P'_{2+}=T^2P_{2+}$. We obtain an approximative attenuated state $\rho'\approx (1-TP_1-T^2P_{2+})\ketbra{0}{0}+TP_1\ketbra{1}{1}+T^2P_{2+}\ketbra{2}{2}$.

The depth of $\rho'$ strongly depends on the choice of the quantum feature. Under the approximation $P_{2+}\ll P_1$, the negativity of the Wigner function demands $T>(2P_1)^{-1}$, which is very challenging to fulfill for single-photon sources. On the other hand, if the NC condition $P_{2+}<\frac{1}{2}P_1^2$ is satisfied before the attenuation, then $P'_{2+}<\frac{1}{2}P'^2_1$ is fulfilled after the attenuation as well. Therefore, the NC withstands any attenuation with $T>0$, which means that the state exhibits infinite NC depth. In contrast, QNG is observable for attenuator transmittances \cite{Lachman}
\begin{equation}
T>\frac{3}{2}\frac{P_{2+}}{P_1^3},\quad{}P_{2+}\ll P_1.
\label{eq.transmittance}
\end{equation}
Note that an arbitrarily small multiphoton contribution $P_{2+}$ makes the QNG depth finite. If $P_1^3$ is substantially larger than $P_{2+}$, the QNG depth can still be very large, even though the state has positive Wigner function. Our goal is to experimentally find such QNG states, which can be very good resources for quantum technology.

{\captionstyle Experimental schemes.---} In our work we used three different systems to generate single-photon states. Of these, two were based on spontaneous parametric down-conversion (SPDC) in a nonlinear crystal. The third system was an InAs/GaAs single quantum dot. The first SPDC source contained a 2-mm-thick type-II BBO crystal that was used in a collinear configuration and was operated in the continuous-wave (cw) regime. Here, the pump power was 90~mW while its wavelength was 405~nm. Correlated photons were spectrally filtered to a bandwidth of 2.7~nm \cite{Jezek}. The second SPDC source produced entangled photon pairs. It contained a 15-mm-long type-II ppKTP nonlinear crystal embedded in a Sagnac-type interferometer loop \cite{predojevic1}. This source was pumped by a 2-ps pulsed laser light of 404~nm wavelength and 80~$\mu$W power per loop direction. The quantum dot sample contained low density self-assembled InAs/GaAs quantum dots embedded in a planar microcavity. The excitation light was derived from a tunable Ti:sapphire laser that could be operated in picosecond-pulsed (82 MHz repetition rate) or continuous-wave mode \cite{predojevic2}. Here, we generated two data sets, one in resonant two-photon excitation using the pulsed mode and the other in above-band continuous-wave mode.

\begin{figure}[b]
	\includegraphics[width=\columnwidth]{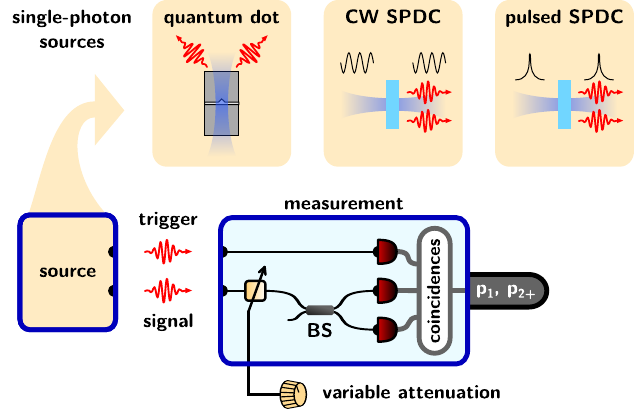}
	\caption{The autocorrelation measurement scheme. The heralded state in the signal arm is attenuated, then $p_1$ and $p_{2+}$ are measured \cite{Jezek}.}
	\label{fig.hbt}
\end{figure}

The measurement scheme was a triggered autocorrelation shown in Fig. \ref{fig.hbt}. Variable attenuation was introduced by moving a blade in the beam. Data acquisition was carried out by a time-tagging module, which stored arrival times of every detection event. The trigger detector conditioned the detections in the signal arm: any detection within a coincidence time window from a trigger detection was considered a coincidence. From these, we measured the probabilities $p_0$, $p_1$, $p_{2+}$, which are estimators of $P'_0,P'_1,P'_{2+}$ \cite{Jezek}. These parameters allowed us to construct the witnesses for NC and QNG states.

{\captionstyle Experimental results.---} In Fig. \ref{fig.mainplot}, we compare the measurement results obtained from all three single-photon sources.
Here, the nonclassicality witness $p_{2+}<\frac{1}{2} p_1^2$ is shown as a solid black line while the QNG witness $p_{2+}<\frac{2}{3} p_1^3$ as a solid blue line. For each single-photon source, we show the results obtained under systematically varied attenuation, given in units of $10\log_{10}(1-T)$ dB. Additionally, we give a theoretical model of the induced losses (dot-dashed lines). These models served us to evaluate the theoretical value of the QNG depth for each source, as given by Eq. (\ref{eq.transmittance}). In addition, we experimentally confirmed the QNG character of the states subjected to a certain maximum attenuation that is the experimentally proven QNG depth. Since it is challenging to experimentally attenuate the state until it is placed exactly on the border of Gaussian mixtures, the proven QNG depth is always lower than the theoretical prediction.

For the pulsed SPDC source (cyan triangles) we estimated the QNG depth to be 14.5 dB and measured 10.8 dB. The cw SPDC source measured with 2-ns coincidence window (orange diamonds) yields a theoretical depth of 19.6 dB and a proven depth of 17.9 dB. Red diamonds stand for weakly pumped cw SPDC. The expected depth is 31.8 dB while the measured value is 18 dB. The state generated by a quantum dot excited above-band (purple squares) shows only nonclassicality and cannot be well compared to SPDC states. With resonant pulsed excitation (blue square), the quantum dot state exhibits QNG character and the theoretical depth is 5.6 dB. Empty blue squares show additional quantum dot states measured with different collection efficiencies. In order to measure the QNG depth for the quantum dot, the measurement time would exceed the stability of the system.

{\captionstyle Measurable depth of NC and QNG.---} As predicted in the discussion above, Fig. \ref{fig.mainplot} shows that the NC depth is robust for all single-photon sources. Both the theoretical models and the directions of the experimental points are parallel to the NC border. In particular, we demonstrated that even with 2 orders of magnitude of attenuation, the data points do not exhibit any trend of approaching the NC border. At very high attenuation, dark counts eventually limit the signal-to-noise ratio. In addition, the required long integration times lead to systematic errors caused by instabilities. Both effects limit the level of attenuation above which no information about the original states can be obtained anymore. This places the NC depth of the single-photon sources beyond measurement. In contrast, the border of QNG can be experimentally reached. Intuitively, this can be understood as follows: the QNG border gives a cubic relation between $p_1$ and $p_{2+}$, while attenuation behaves quadratically. This makes the QNG depth a measurable feature for single-photon states, and, consequently, a convenient benchmark for single-photon sources. Furthermore, QNG shows high robustness for SPDC sources, which proves that the generated states can be considered high-quality single-photon states defined in Eq. (\ref{eq.hqstate}).

{\captionstyle Experimental optimization of SPDC sources.---} Optimization of the QNG depth can be achieved via certain experimental parameters, depending on the source. These parameters include pump power, the width of the coincidence window, SPDC efficiency and losses. Inevitable losses in the experimental setup decrease the QNG depth for all types of sources. Optimization of the coupling or collection efficiencies is therefore essential, as well as high quantum efficiency of the detectors.

The impact of the other parameters on the QNG depth is not straightforward to see. Previously, the effects of some experimental factors were examined using a QNG witness quantification $\Delta W$ \cite{Jezek}. The witness, however, intrinsically differs from the QNG depth.

Generally, QNG depth increases with lower pump power and conversion efficiency. For the cw pump, the coincidence window has an optimum width depending on the detector time resolution. For the pulsed pump, the coincidence window is upper bounded by repetition rate and lower bounded by the photon lifetime and detector time resolution. Moreover, when considering a cw source and a comparable pulsed source, the cw source intrinsically yields higher QNG depth. More detailed discussion can be found in the Appendix \cite{appendix}.

{\captionstyle The quantum dot source.---} Quantum dots generate fundamentally different states of light than SPDC. They rely on formation of an electron-hole pair and subsequent recombination that results in photon emission. Specifically, the recombination of the biexciton gives rise to two spectrally distinct photons emitted in a time-ordered cascade. In our measurements, the first photon of the cascade serves as a trigger for the second photon. If we consider resonant excitation by a picosecond laser, only the transition between the vacuum state and a single biexciton is possible. The decay time of the biexciton is 2 orders of magnitude longer than the pump pulse. Therefore, the probability to systematically generate a multiphoton state by a single pulse is very low. This is an extremely valuable asset, which potentially makes quantum dots much closer to an ideal single-photon source than SPDC.

In practice, however, there is always some background noise present in the measurement, that is responsible for the $p_{2+}$ contribution. The quantum dot state in Fig.~\ref{fig.mainplot} (blue square) shows that this noise is stronger than the noise of an attenuated SPDC single photon (red diamond) operated in the cw regime. The QNG depth can be improved by increasing the collection efficiency \cite{Senellart1,Senellart2}. As a result, one can expect an increase in $p_1$ with $p_{2+}$ remaining constant. The three blue-square points in Fig. \ref{fig.mainplot} show measured states with various degrees of efficiency. If the collection efficiency improves by a factor of 9, the quantum dot would yield states with higher QNG depth than the cw SPDC state. The results presented in Ref. \cite{Senellart1} indicate that by embedding the quantum dot in a micropillar cavity, one can reach a factor of 16 improvement. For such collection efficiency, the QNG depth may exceed 40 dB and surpass the QNG depth of the SPDC.

{\captionstyle Conclusion.---} We experimentally verified high QNG depths of various single-photon states. This is in strong contrast to the fragility of the Wigner function negativity; therefore, our results demonstrate that QNG is a robust quantum resource.

It can be seen that SPDC produces single-photon states with extremely robust QNG depth. The data prove that with commonly used single-photon sources, quantum non-Gaussianity can be preserved after propagating the photon through 8 km of fiber, assuming 4 dB/km losses for the wavelength of 0.8 $\mu$m. For similar sources at telecom wavelength, the range is about 180 km.

When compared to a quantum dot, SPDC can generate much more robust states at present, but its noise is fundamentally unavoidable. Further improvement of the technical aspects of quantum dot sources could lead to single-photon states more robust than those generated by SPDC.

This research has been supported by the Czech Science Foundation (13-20319S). The research leading to these results has received funding from the European Union Seventh Framework Programme under Grant Agreement No. 308803 (project BRISQ2). L.L. thanks the Czech-Japan bilateral project LH13248 of the Ministry of Education, Youth, and Sports of Czech Republic. M. Mikov\' a acknowledges the support of Palack\' y University (IGA-P\v rF-2014-008). A.P. acknowledges the support of the University of Innsbruck, given through Young Researcher Award. Additionally, the work at the University of Innsbruck was partially supported by the European Research Council, project "EnSeNa" (257531). G.S.S. acknowledges partial support through the NSF Physics Frontier Center at the Joint Quantum Institute (PFC@JQI).

\section{APPENDIX}

\subsection{The effects of experimental parameters}

The two SPDC sources we used were operated in different regimes -- with pulsed and CW pumping. Nevertheless, the photodistribution of the generated two-mode state before losses is the same: $\mathcal{P}(m,n)=\delta_{mn}(1-g)g^n$. The temporal width of the state is given by the coincidence window $\tau$. $\mathcal{P}(m,n)$ is the probability of generating $m$ and $n$ photons in the two respective modes, $\delta_{mn}$ is the Kronecker delta and $g$ is the gain, which is proportional to pump power and SPDC efficiency. The gain is responsible for the systematically generated component in the heralded single-photon state, since $p_{2+} \propto g$ for $g \ll 1$. Since the QNG condition dictates $p_{2+}<\frac{2}{3}p_1^3$ and the depth is thus estimated as $T_{\text{min}}=\frac{3}{2}\frac{p_{2+}}{p_1^3}$, $g$ needs to be minimal. It is usually well under experimental control, but cannot be decreased arbitrarily due to practical limitations.

For typical CW-pumped sources, the detected state is temporally multi-mode, which effectively means $p_{2+} \propto \tau$, where $\tau$ is much larger than the temporal bandwidth of the biphotons. It follows that the coincidence window needs to be optimized, too. Its width is lower-bounded by the resolution time of the detectors. If the coincidence window is reduced below that limit, it effectively introduces a loss to the state and decreases the QNG depth. On the other hand, if the coincidence window is excessively large, the higher $p_{2+}$ contribution decreases the QNG depth as well. Therefore, there is an optimum coincidence window that maximizes the QNG depth.

The coincidence window plays a minor role in the pulsed regime. There, it has no effect on $p_{2+}$, assuming the coincidence window is not shorter than the lifetime of the photons and not longer than the delay between pulses. Thus, there is a fixed pump energy contributing in each coincidence window. Analogously to the CW regime, if the coincidence window is shorter than the detector resolution time, the effective loss decreases the QNG depth. Therefore, as long as the coincidence window remains within the aforementioned limits, it has no effect on the QNG depth.

\subsection{Comparing CW and pulsed pumping regime for SPDC}

Let us consider two cases: a CW and a pulsed single-photon source with an identical frequency of heralded state generation -- a heralding rate. Let both sources have identical average pump power $\bar{S}$, the same overall conversion efficiency, number of modes and effective losses in the set-up. In the low-gain approximation $g \ll 1$, such two sources would have similar heralding rate and $p_1$, but different $p_{2+}$. Namely, the $p_{2+}^{\text{CW}} \approx \mu \bar{S} \tau$ and $p_{2+}^{\text{pul}} \approx \mu \frac{\bar{S}}{\nu}$, where $\nu$ is the repetition rate of the pulsed pump, $\tau$ is the width of the coincidence window, and $\mu$ a common proportionality constant. Since the QNG depth of the heralded state is given by $T_{\text{min}}=\frac{3}{2}\frac{p_{2+}}{p_1^3}$, the ratio of the minimum transmittances $T^{\text{CW}}_{\text{min}} / T^{\text{pul}}_{\text{min}} \approx \tau \nu$. The coincidence window can be minimized to the limit of the detector resolution, typically $\mathop{\sim}10^{-9}$ s. The repetition rate of the pump laser is not a very flexible parameter ($\mathop{\sim}10^{7}$ s$^{-1}$) and cannot be routinely adjusted. That gives a significant difference in the QNG depth in favor of CW-pumped SPDC sources, assuming similar heralding rates. Furthermore, the CW source often has significantly more modes than the pulsed, which can effectively lead to an additional factor of $\sim\frac{1}{2}$ in $T^{\text{CW}}_{\text{min}} / T^{\text{pul}}_{\text{min}}$. There are ways for pulsed sources to effectively increase the repetition rate and decrease the $p_{2+}$ contribution \cite{White}. However, for repetition rates approaching detector resolution, the pulsed source would approach the CW source, but would never yield a larger QNG depth.

\end{document}